\documentclass[fleqn,10pt]{wlscirep}
\usepackage[utf8]{inputenc}
\usepackage[T1]{fontenc}
\usepackage[sort&compress,numbers]{natbib}

\usepackage{cleveref}

\usepackage{xr}
\title{Team careers in science: formation, composition and success of persistent collaborations}

\author[1]{Sandeep Chowdhary}
\author[1]{Luca Gallo}
\author[2]{Federico Musciotto}
\author[1,*]{Federico Battiston}

\affil[1]{Department of Network and Data Science, Central European University, Vienna, Austria}
\affil[2]{Department of Physics and Chemistry, University of Palermo, I-90128 Palermo, Italy}

\affil[*]{battistonf@ceu.edu}

\newcommand{\san}[1]{\textcolor{blue}{#1}}

\begin{abstract}
Teams are the fundamental units propelling innovation and advancing modern science.
A rich literature links the fundamental features of teams, such as their size and diversity, to academic success. 
However, such analyses fail to capture temporal patterns, treating each group of co-authors as a distinct unit and neglecting the existence of persistent collaborations. 
By contrast, teams are dynamical entities, made of core members who consistently work together, surrounded by transient members who sporadically participate. 
Leveraging on a large dataset of over 205 million scientific papers published since 1900, we extract $511,550$ core teams of statistically significant persistent collaborations of pairs and larger groups of scientists. 
We look into `team careers' investigating their trajectories in time, characterizing their formation, productivity and eventual dissolution. 
We characterize team composition along multiple dimensions, including age, academic affiliation and scientific disciplines.
Finally, we investigate the academic impact of persistent collaborations, hallmarking the key compositional features underlying their success.
Our work sheds light on the nature of persistent teams, informing researchers, institutions and funding agencies about the dynamics of their formation, evolution and success.
\end{abstract}

\begin{document}

\flushbottom
\maketitle

\thispagestyle{empty}

\section*{Introduction}
The explosion in the amount of data on scientific research has recently offered extraordinary opportunities to investigate the evolution of science.
The availability of multiple, massive data-sources, along with the collaborative and interdisciplinary effort of scientometricians, natural, computational and social scientists, has led to the emergence of the interdisciplinary field of Science of science \cite{hicks2015bibliometrics, zeng2017science, clauset2017data, sugimoto2018measuring,fortunato2018science, liu2023data}, which aims at quantifying the fundamental mechanisms underlying patterns and behaviors in scientific research. 
Within this field, over the last decade a large body of literature has focused on the unfolding of individual scientists' careers.
The temporal patterns of productivity and impact of researchers were investigated in detail, revealing that the most-cited work of a scientist occurs randomly within their career \cite{sinatra2016quantifying}, and that their high-impact papers are likely to come in close succession \cite{liu2018hot,liu2021understanding}.
Further research explored how the career of scientists is affected by their individual characteristics, such as gender \cite{lariviere2013bibliometrics,huang2020historical} and ethnicity \cite{bertolero2020racial}, as well as their academic choices and opportunities, including being affiliated with certain institutions \cite{Murray2016,zhang2022labor,krauss2023early}, moving to a different one \cite{deville2014career}, switching research topic \cite{zeng2019increasing}, or dropping out of academia \cite{milojevic2018changing}.

In modern science, research is rarely conducted in isolation \cite{bozeman2014research} and the academic trajectory of individual scientists is thus deeply influenced by their collaborations.
At the early stage of the career, co-authoring a paper with a top scientist can guarantee a competitive advantage that results in becoming a top scientist as well \cite{li2019early,krauss2023early}.
Being mentored by a highly prolific researcher can determine future success, in terms of impact, academic proliferation and job opportunities \cite{malmgren2010role,lienard2018intellectual}.
Senior colleagues can also transfer to younger researchers the expertise needed to publish in prestigious journals, which goes with higher citation rates \cite{sekara2018chaperone}.
Moreover, junior researchers who collaborate with scientists from other continents are more likely to secure funding from public institutions \cite{chowdhary2023dependency, chowdhary2023funding}.
Conversely, the sudden end of collaborations with prominent colleagues can have a negative impact on productivity \cite{azoulay2010superstar,borjas2015peers}.

While in the past centuries lone geniuses had played a decisive role in science, in more recent times scientific revolutions are more likely to emerge from the work of teams \cite{simonton2013scientific}.
In 1930s, the \textit{Via Panisperna boys} led by Enrico Fermi discovered slow neutrons, paving the way to the development of nuclear reactors \cite{segre2016pope}. 
Between 1947 and 1948 at Bell's Labs, John Bardeen, Walter Brattain and William Shockley invented the first working transitors, the ``nerve cells'' of the Information Age \cite{riordan1999invention}.
In 2012, a team led by Jennifer Doudna and Emmanuelle Charpentier developed the CRISPR–Cas9 gene-editing tools, pioneering a revolution in molecular biology \cite{ledford2020pioneers}.
At a larger scale, long-lasting international collaborations such as LIGO and VIRGO \cite{abbott2016observation}, CERN \cite{aad2012observation,chatrchyan2012observation} and the Human Genome project \cite{international2001initial} expanded the frontiers of human knowledge.
In general, research is progressively dominated by teams of scientists, as they are not only more productive than individual authors but also more impactful \cite{guimera2005team,wuchty2007increasing} and innovative \cite{uzzi2013atypical}.
The increasing predominance of teams has been linked to the need of gathering together in interdisciplinary collaborations researchers with specialized expertise to solve modern-day scientific problems \cite{hunter2008collaborative,jones2008multi}.
Inevitably, understanding the determinants of teams success in academia has become a crucial area of investigation. 
For instance, recent works showed that team size plays an important role \cite{wuchty2007increasing}, with smaller teams often producing more innovative work\cite{wu2019large}.
Similarly, team diversity, in terms of geography and affiliation \cite{jones2008multi,hsiehchen2015multinational,coccia2016evolution,gazni2012mapping}, but also ethnicity \cite{alshebli2018preeminence}, and gender \cite{nielsen2017gender,yang2022gender}, also significantly affect scientific impact.

Despite the vast literature on team science, understanding the temporal patterns of scientific collaborations has so far been largely overlooked. 
Akin to those of individual scientists, some scientific teams follow consistent career trajectories, the features of which remain largely unknown.
Persistent collaborations among ``science buddies'' \cite{yanai2024takes} are indeed considered pivotal to scientific research \cite{hoy2019science}, positively contributing to productivity and impact \cite{petersen2015quantifying,liu2024understanding}. 
While team formation and evolution has garnered some attention \cite{guimera2005team, milojevic2010modes,milojevic2014principles}, most research treats each group of co-authors as a distinct unit.
The few works accounting for the temporal characteristics of collaborations are limited to pairs of scientists \cite{son2023untangling}, or focus on their impact on individuals researchers \cite{petersen2015quantifying}.
Very recently, it was revealed that teams of researchers who have not previously worked together tend to publish more original and multidisciplinary successful papers \cite{zeng2021fresh}.
However, at large, unraveling the temporal dimensions of teams in science represents an unexplored territory.

Here, we shift the focus and reveal the determinants of career trajectories of persistent scientific collaborations.
Exploiting a large dataset of over 205 million research articles published since 1900, we build a vast hypergraph of collaborations connecting more than 90 million scientists across various disciplines, where hyperedges encode groups of co-authors who published together. 
Leveraging higher-order network theory~\cite{battiston2020networks}, we use a statistically-validated approach\cite{musciotto2021detecting, musciotto2022identifying} to reveal and extract $511,550$ cores of persistent collaborators.
We explore the temporal patterns behind their formation, activity and eventual dissolution. 
We characterize the composition of persistent cores along multiple dimensions, including age composition, geographical diversity and disciplinary expertise, as well as the role of transient team members.
We conclude by investigating the temporal and compositional key features underlying the academic success of persistent teams.

\section*{Results} 

\begin{figure}[t]
\centering
\includegraphics[width=.8\textwidth,trim={0cm 0cm 0cm 0cm},clip]{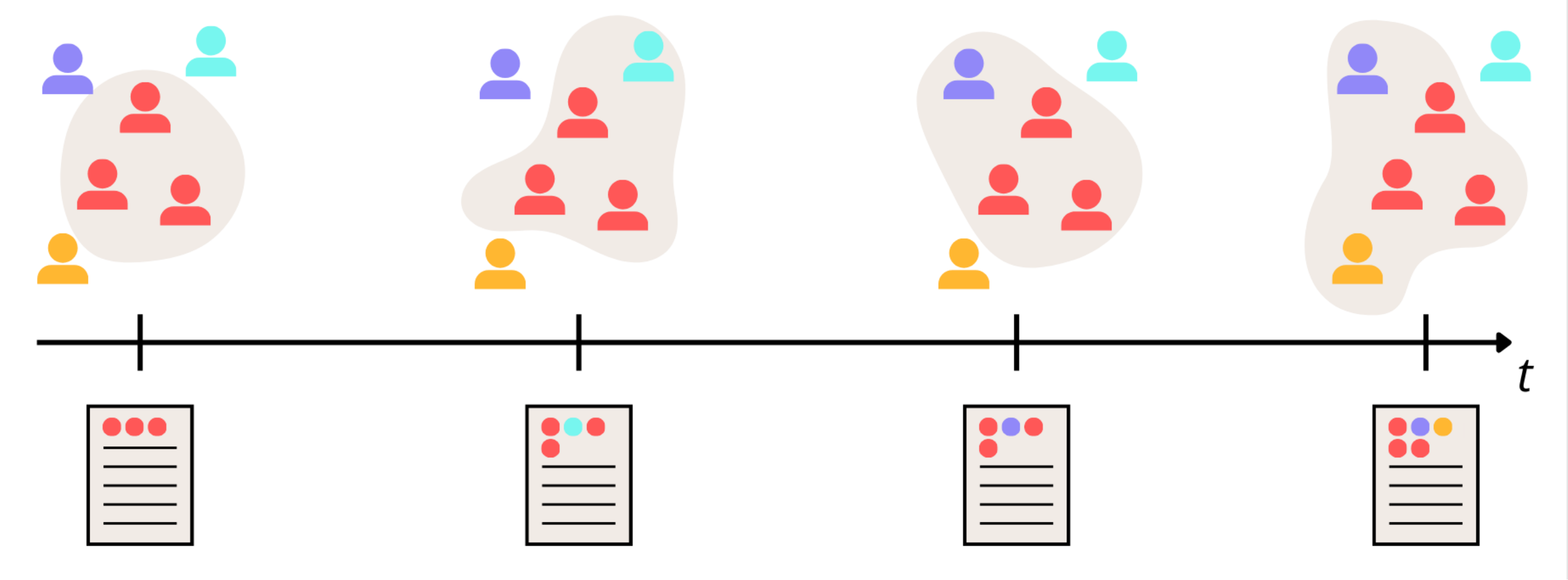}
\caption{\textbf{Identification of cores of persistent collaborators}. 
We build a temporal hypergraph encoding scientific collaborations, where each set of authors of a scientific paper is represented as a hyperedge (grey-shaded areas; see Methods for details).
Typically persistent core members (red) who consistently collaborate do not work in isolation, but often publish together as part of larger teams which also include transient members (other colors).
We identify sets of statistically significant collaborators, and investigate the temporal careers of such persistent cores. 
}
\label{fig:fig0}
\end{figure}

\textbf{Formation, productivity and dissolution of persistent scientific collaborations.} 
Not all members of a scientific team collaborate in the same way. In science, a team is often composed of core members, who work persistently and recurrently together over the years, surrounded by transient members, who come and go in the collaboration.
Since the exact set of co-authors can change from paper to paper, this makes following the trajectories of teams much more complicated than the careers of single individuals.
To study team careers, our first step is thus to identify persistent scientific collaborations from empirical data.
We analyze a large dataset collected from OpenAlex \cite{openalex,chawla2022massive}, consisting of 248 million journal papers published since 1900, covering 90 million scientists across various scientific disciplines.
From the publication records, we build a hypergraph~\cite{battiston2020networks} of scientific collaborations, where each hyperedge encodes the set of co-authors of a paper, and use a statistically-validated approach \cite{musciotto2021detecting, musciotto2022identifying} to extract those groups of scientists that have persistently published together over their careers (Fig.~\ref{fig:fig0}, see the Methods for a detailed description of the data and the methodology).
Through this procedure, we identify $511,550$ persistent scientific collaborations. 

\begin{figure*}[ht]
\includegraphics[width=1\textwidth,trim={0cm 0cm 0cm 0cm},clip]{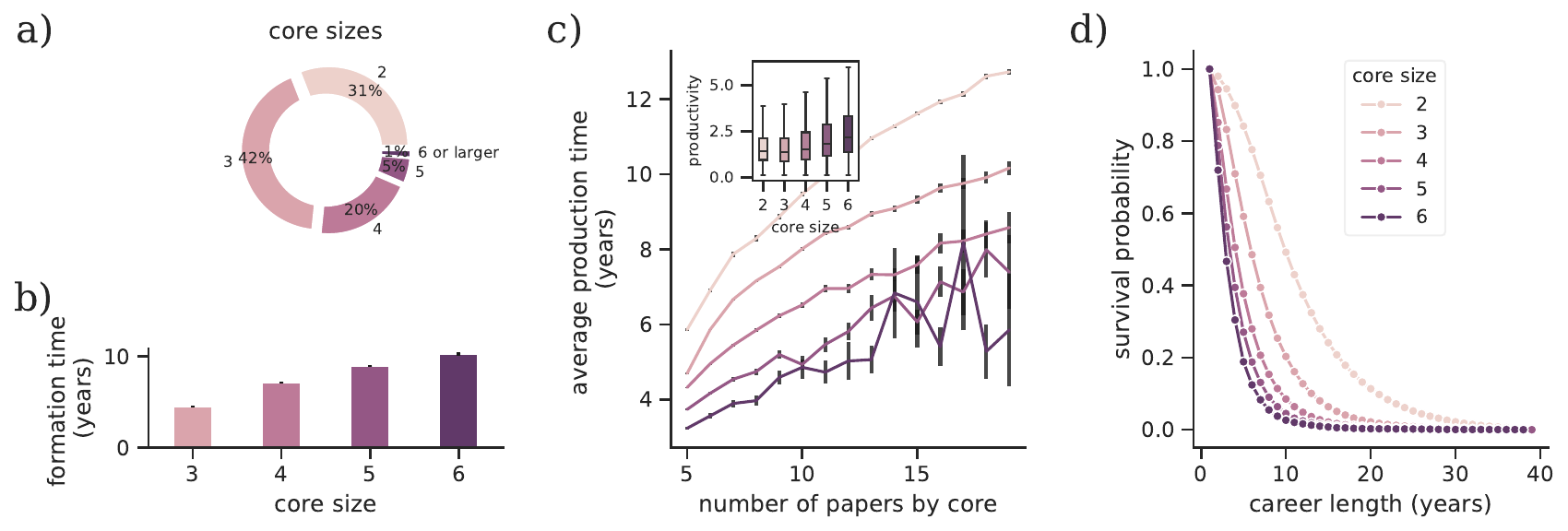}
\caption{\textbf{Formation, productivity and dissolution of persistent team cores.}  
(a) Distribution of the number of scientists per team core. 
(b) Formation time of cores as a function of core size.
(c) Production time for a given number of papers for cores of different sizes. 
Inset: average number of papers published per year as a function of the core size. (d) Survival probability of a core as a function of the career length, i.e., the time since its formation. }
\label{fig:fig1}
\end{figure*}

We begin by investigating the typical number of members in persistent scientific teams. 
We compute the distribution of core sizes (Fig.~\ref{fig:fig1}a), finding that the greatest percentage of cores ($42\%)$ have 3 members, followed by cores of size 2 ($31\%$) and 4 ($20\%$). 
Larger persistent collaborations are rarer, with the fraction of cores of size 7 or higher being less than 1 in 100.

Typically, the formation of a scientific team requires a significant amount of time. 
Indeed, a persistent team may start as a small number of scientists working together, and gather further members around it later on.
We thus ask: How fast do cores assemble?  
To examine this, we evaluate the time elapsing from the first publication authored by any subgroup of core members to the first publication authored by all members of the team.
We refer to this quantity as the formation time.
Note that, by our definition, teams of two members are established with a formation time of zero.
The average formation time for cores of different sizes is shown in Fig.~\ref{fig:fig1}b. 
We observe that smaller cores gather faster than bigger ones.
In particular, teams of three members typically take 4.6 years to form, 
while larger cores take more and more time to gather, i.e., 7.3 years for cores of size 4, 9.2 years for cores of size 5, and 10.4 years for cores of size six. 
Furthermore, we note that a certain percentage of cores are formed instantaneously, i.e., they have a formation time of zero, meaning that the first publication by the team includes all of its core members.
Out of all cores of size 3, nearly $16.8\%$ assembled instantaneously. 
For larger cores, this number drops, $5\%$ (for size 4), $2.3\%$ (size 6), and only $1.4\%$ (for size 6).

After their formation, teams start collaborating and publishing about their research. 
While the efficiency of scientific teams is hard to quantify, as we have no information on how long a team worked on a publication, we can analyze their career in terms of the number of joint publications as a function of time.  
To this end, we consider all cores of a given size that have published a certain number of scientific papers and measure the average time taken to produce them (Fig.~\ref{fig:fig1}c). 
Moreover, we compute the yearly-production-rate, i.e., the productivity, for different core sizes (inset). 
Our analysis reveals that bigger cores outproduce smaller cores, taking less time on average to publish the same number of scientific articles. 
Such a result calls for an in-depth analysis of how persistent scientific teams communicate, coordinate and organize as a function of the number of their members \cite{hall2018science}.
Our observation may also partly explain the increasing dominance of teams in the production of science and arts \cite{wuchty2007increasing}.

Though they may persist for a long time, all scientific collaborations eventually end.
Therefore, we examine the typical lifespans of persistent scientific teams. 
We calculate the survival probability of cores as a function of the career length, i.e., the time elapsed from their first to their last publication. 
We find that smaller cores are typically more persistent (Fig.~\ref{fig:fig1}d). 
For instance, nearly $45\%$ of scientist-duos work together for at least 5 years, while some of them can keep working together for as long as 30 years after their first joint publication.
Larger cores, instead, have shorter careers: It is highly atypical for team cores of 4 or more scientists to continue publishing together after 10 years since their first publication.  

\begin{figure*}[ht]
\includegraphics[width=1\textwidth,trim={0cm 0cm 0cm 0cm},clip]{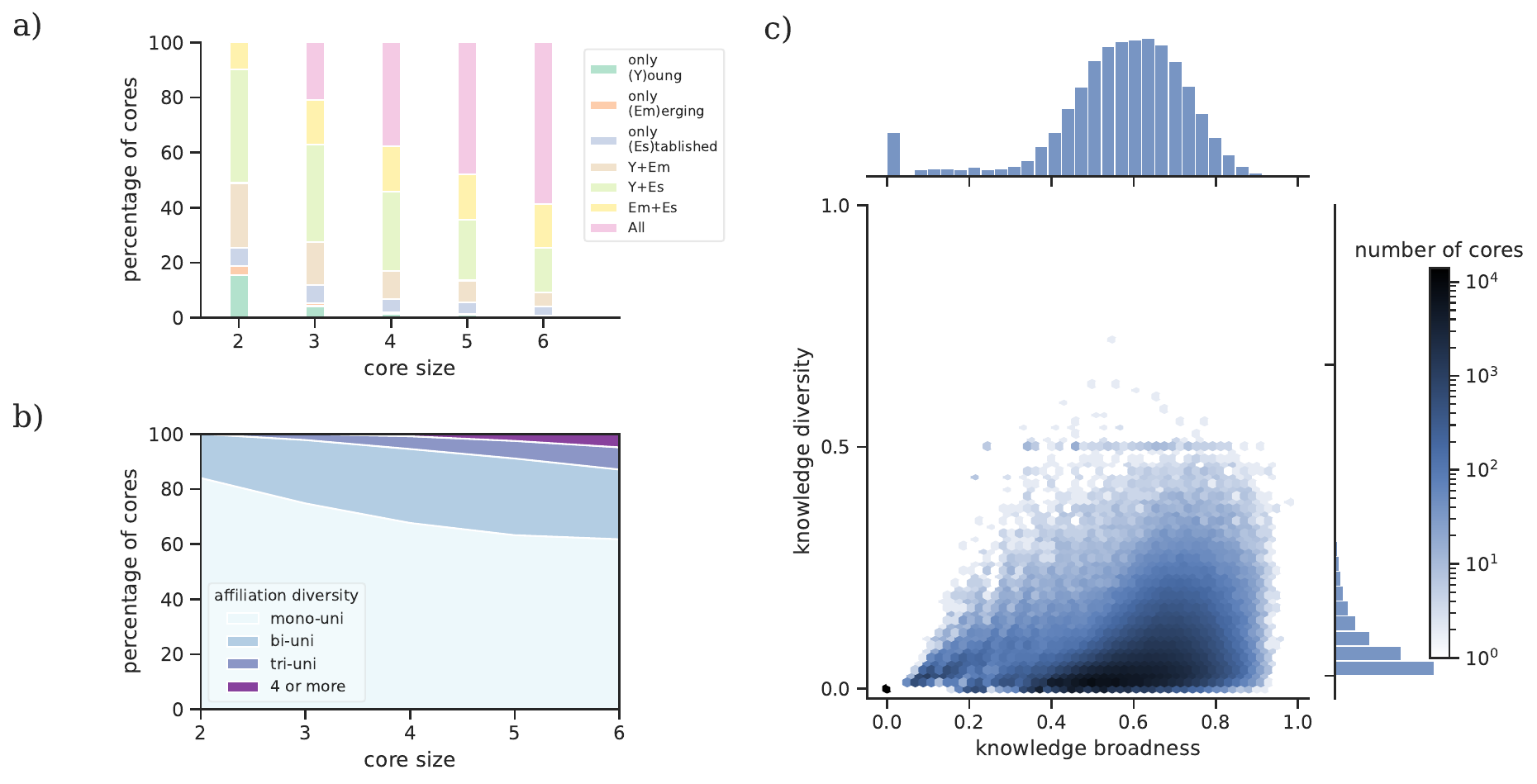}
\caption{\textbf{Composition of persistent scientific cores.}   (a) Percentage of each possible age composition of the core as a function of core size. (b) Percentages of cores that are spread across 1 (mono), 2 (bi), 3 (tri), and 4 or more universities as a function of core size. (c) Interdisciplinary diversity. The joint distribution knowledge diversity (quantifying cross-member conceptual distances) and knowledge broadness of the team (entropy of the sum of individual concept vectors of members). For monodisciplinary cores where all members belong to the same field both knowledge broadness and core diversity are zero.}
\label{fig:fig2}
\end{figure*}

\textbf{Composition of persistent collaborations.} The members of a persistent collaboration can be identified by various characteristics, including age, affiliations with universities, and scientific expertise.
Understanding the composition of persistent teams in terms of the individual characteristics of their members, i.e., whether they overlap, match, or integrate, can shed light on the mechanisms that facilitate long-lasting collaborations.
For instance, a persistent team may show age-homophily among its members or, instead, it may comprise young researchers and older experienced scientists, e.g., a long-standing collaboration between mentor and mentee. 
To understand the role played by age in persistent collaboration, we compute the career age of each scientist in the team (i.e., the time passed since the scientist's first publication), which we then use to characterize the age-composition of persistent teams at the time of core creation (i.e., the first joint publication). 

\begin{figure*}[ht]
\includegraphics[width=1\textwidth,trim={0cm 0cm 0cm 0cm},clip]{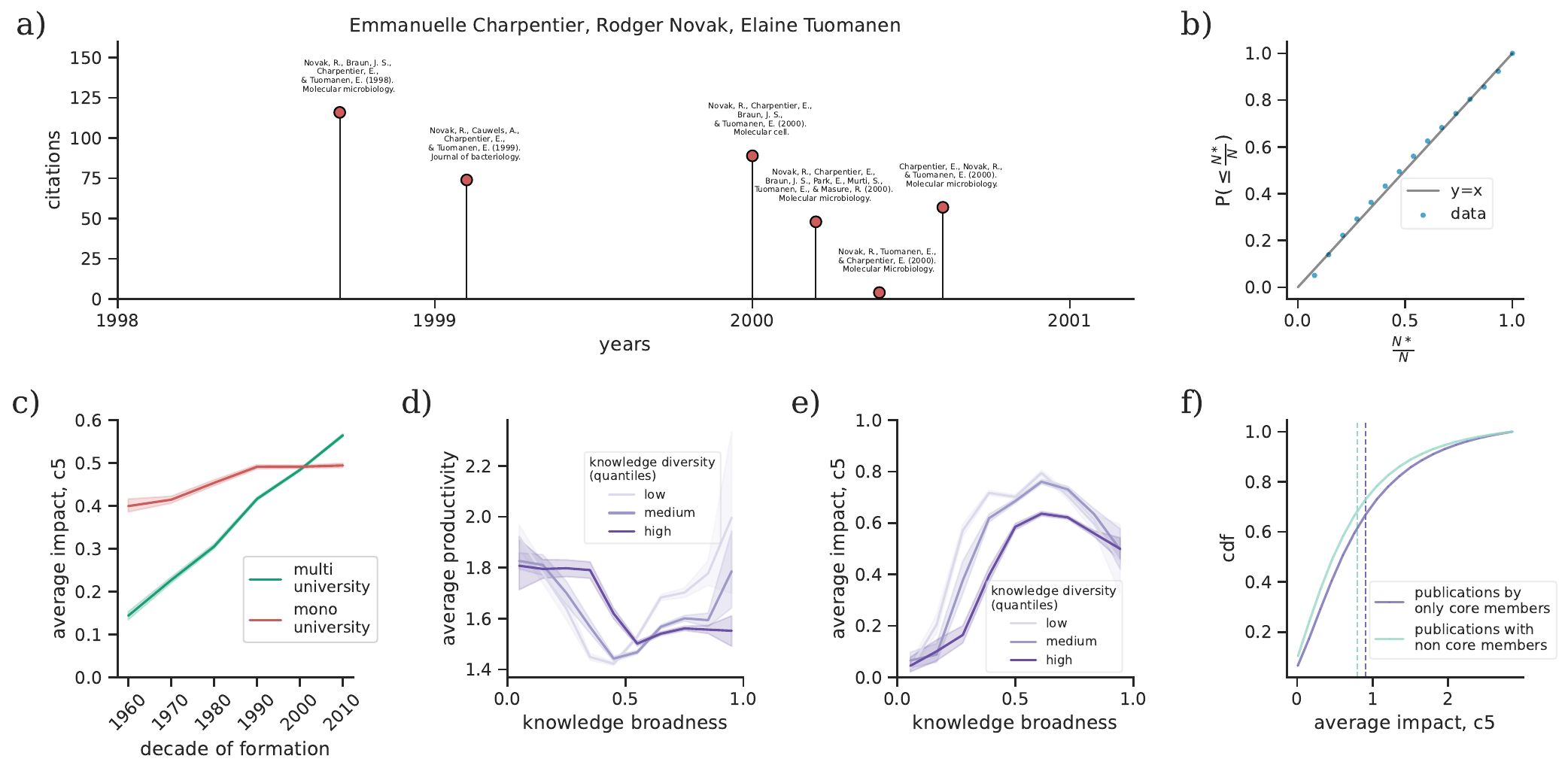}
\caption{\textbf{Temporal dynamics and composition-based correlates of team success.} (a) Number of citations received by publications of a persistent collaboration involving the Nobel laureate Emmanuelle Charpentier (Nobel Prize in Chemistry, 2020). (b) Cumulative distribution P($\leq$N*/N) for cores with joint total papers (N)$\geq$10, where N*/N denotes the order of the highest-impact paper in a core’s career, varying between 1/N and 1. The cumulative distribution of N*/N is almost a straight line with slope $\approx$1, indicating that N* has roughly the same probability to occur anywhere in the sequence of papers published by a core. (c) The average impact of cores that formed in a given decade, where cores are separated into those co-located at the same university, and others with multiple institutions. (d) Average productivity as a function of knowledge broadness of the core. 
Cores were also assigned categories according to core diversity into 3 equal-sized categories --low, medium or high. (e) Average impact as a function of knowledge broadness of the core. 
Cores were also assigned categories according to core diversity into 3 equal-sized categories --low, medium or high.
(f) Cumulative probability distribution of average impact of team publications authored solely by the core members (blue) or involving transient, non-core members (green).}
\label{fig:fig3}
\end{figure*}

We divide scientist into three age groups, namely young ("Y", career age less than 7 years), emerging ("Em", between 7 to 14 years) and established scientists ("Es", more than 14 years).  
We thus assign each core to one of 7 possible categories based on the age groups of the members.
We distinguish cores where all scientists belong to the same age group (only young, only emerging, or only established), cores where two age groups are represented (young + emerging, young + established, or emerging + established), and cores with scientists from all three groups. 
The percentage of each category in the data is shown for various core sizes in Fig.~\ref{fig:fig2}a. 
For dyadic cores, the most predominant age composition at time of assembly is a young and an established scientists working together, a typical Student-Professor motif. 
This is followed in frequency by the young+emerging motif.  
For triadic cores, the mixed age composition i.e. Y+Em+Es likely corresponding to the typical PhD-PostDoc-PI core, starts to be a prominent age composition. 
Still, the dominant composition is of only Young + Established researchers. 
Beyond triads, for core size 4 and higher, mixed cores become increasingly common. 
We also analyze how the formation time and career length of a core vary as a function of its age composition, revealing shorter assembly times and longer careers for teams featuring younger members (Supplementary Fig. S1 and S2).

Next, we characterize how diverse persistent collaborations are in terms of academic affiliation. 
We will refer to this as the affiliation diversity of the cores. 
To quantify it, for each paper of the core, we consider the set of affiliations that the members have at the time of publication, and evaluate the minimum number of affiliations needed to represent all members of the core. 
For illustration, consider a core of 3 members, A, B, and C, having the affiliations A:[MIT, UCSD], B:[MIT, UCSD], C:[MIT]. 
In this case, the minimum number of affiliations needed to represent the core is 1, as all members share one affiliation (MIT).
Instead, if we consider the case A:[Caltech, UCSD], B:[UCSD, Indiana], C:[Caltech], we would need at least 2 affiliations (Caltech and Indiana) to cover all members of the core.
As the core members can change academic affiliation during the lifetime of the collaboration, we evaluate the minimum number of affiliations for each article published and define the affiliation diversity of the core as the most common value.
Fig.~\ref{fig:fig2}b shows the fraction of cores of different sizes being covered by one, two, or three universities. 
Possibly surprisingly and in contrast with an increasing trend to collaborate remotely~\cite{teasley2001scientific}, we note that 75.8$\%$ of cores are situated at the same university, hinting at the importance that common institutions play in sustaining long-term collaboration. 
It is also worth noting that cores that are not based at the same university are usually located at 2 universities (21.7$\%$), while only very rarely persistent cores span 3 or more institutions (2.5$\%$). 
Geographically speaking, $78.1\%$ ($84.7\%$) of persistent cores are situated within the same country (continent). 
We also find that co-presence at the same institution is associated with a shorter formation time and a longer lifespan of the core (Supplementary Fig. 1, 2).

Finally, we aim to comprehend how diverse are core members in terms of their scientific expertise, namely whether the teams are mono-disciplinary or interdisciplinary.
To achieve a data-driven understanding of the extent to which topic composition affects persistent collaborations, we measure two complementary dimensions of team interdisciplinarity, namely knowledge broadness and knowledge diversity.
Knowledge broadness captures the breadth of the combined expertise of the persistent core. knowledge diversity, by contrast, quantifies how much team members are diverse with regards to the disciplines they are associated with.

Both measures range from 0 to 1, where 1 are achieved for maximum values of multidisciplinarity of the team (broadness) and topic complementarity across members (diversity, see Methods for details).

Fig.~\ref{fig:fig2}c shows the joint distribution of knowledge broadness and knowledge diversity across the cores.
We observe that only a small fraction of teams, almost $3\%$, are completely mono-disciplinary, namely all members work in one scientific field.
Also, we notice that the density of cores tapers off as diversity increases, with few cores having a diversity larger than 0.5.
This suggests that the core members should have a minimum amount of disciplinary overlap for their collaborations to persist. 
The vast majority of cores, however, are broad in their knowledge base (over $50\%$ have more than $0.6$). 
In particular, a significant proportion of cores ($\sim 9.1\%$) display broadness larger than $0.75$, indicating teams which work at the interface of several disciplinary fields. 
Yet, knowledge diversity never gets close to its maximum value, which would correspond to a team where 
no shared topics exist between its members. 
This highlights the importance of topic overlap to sustain persistent collaborations. 
Moreover, we find that teams with low knowledge diversity across members take less time to form and have longer careers, further supporting the association between team persistence and topic synergy, whereas low knowledge broadness, on the other hand, is associated with shorter lifespans and a slower formation process (Supplementary Fig. 1,2).

In the SI we discuss team exclusivity, namely the tendency of the members to publish exclusively with the core, and provide an analysis of the demographics and contribution of the transient members surrounding cores of persistent collaborators.

\textbf{Team success.} 
Peak performance in individual scientific careers is known to be randomly distributed, a result known as the random impact rule \cite{sinatra2016quantifying}. 
In other words, the most-cited article in a scientist's career can be, with an equal probability, any paper they published, from the first to the very last publication. 
Analyzing the publication history of persistent scientific collaborations (Fig.~\ref{fig:fig3}a), we aim to understand the pattern of success in team career, and how their composition affects their impact.
To examine whether team careers also exhibit the random impact rule, we adopt a similar methodology as \cite{sinatra2016quantifying,janosov2020success} and measure the relative position N* of the highest-impact paper in a core's career, i.e., in the sequence of its N publications.
We measure the impact as the number of citations after five years (c5), normalized to account for inflation and large variations between different disciplines \cite{radicchi2008universality}.
Fig.~\ref{fig:fig3}b shows the cumulative distribution function $P(\leq N^*/N)$, namely the probability that the most-cited work of a core appears before the N*-th publication. 
We observe that the function nearly follows the cumulative probability of a uniform distribution, which would indicate that the highest-impact article of a core can be published at any point in the career.
While the most-cited article can occur at any time, the average impact of a team throughout its career can show non-random patterns.
In particular, our analysis captures a higher average impact in the first half of the career compared to the second half, hinting that freshness in an early career may bring a higher impact (Supplementary Fig. 5).
Our findings are in agreement with results showing how the team freshness has a positive effect on the team impact \cite{zeng2021fresh}. 
Another universal feature of individual creative careers is the presence of hot-streaks, periods of clustered high-performance works observed in various contexts, from science \cite{liu2018hot, liu2021understanding} to arts and other creative domains\cite{williams2019quantifying}. 
Our results show a tendency for highest-impact publications to be clustered, i.e., to occur close to each other in the team's career, thus confirming the presence of hot-streaks (Supplementary Fig. 6).

Having characterized the diversity of persistent collaborations in terms of age, affiliations, and scientific expertise, we are now interested in understanding how those features contribute to their academic success. 
First, we measure the average paper impact, i.e., the average normalized number of citations after five years, for all possible core age compositions in our data. 
Cores composed of members from the same age groups perform typically worse than collaborations with scientists from multiple age groups (Supplementary Fig. 7). Besides the age composition, the working location of a core team can affect the impact of the work it produces. 
To test this, we compare the average paper impact after five years for mono and multi-university cores. 
In agreement with previous research \cite{jones2008multi}, at the aggregated level we observe that multi-university teams are more successful than mono-university ones (Supplementary Fig. 8).
However, a time-resolved analysis of the average impact, consisting in comparing mono and multi-university cores as a function of the year of formation, reveals a more nuanced picture (Fig.~\ref{fig:fig3}c). 
We find that the impact advantage of multi-university core is a recent phenomenon.
Mono-university cores formed before the 2000s have on average a higher impact compared to multi-university collaborations formed during the same years.
However, while more recently formed mono-university cores do not show a significantly higher impact compared to older collaborations, multi-university cores formed in the last decade have almost three times the impact of cores started in the 60s.
Hence, multi-university cores formed after year 2000 are more successful than mono-university cores formed in the same period.
A possible explanation for this observation lies in recent technological progress, from the development of computer-based mailing systems to the advent of internet and other communication technologies, which have enhanced the experience of remote collaboration, limiting the logistic advantage associated with working in close physical proximity.

In addition to impact, we look at the productivity of teams as a function of knowledge broadness, grouping the cores based on the level of knowledge diversity, i.e., low, medium, and high diversity, respectively (Fig.~\ref{fig:fig3}d). 
We find that the relationship between productivity and knowledge broadness approximately follows a U shape. 
This suggests that teams with a focused approach are more productive than those with moderate levels of diversity. 
However, productivity increases once again for teams exhibiting highest degrees of broadness. 
For a given broadness, teams high in knowledge diversity show higher productivity.
As an additional analysis, we study how productivity depends on the core composition in terms of age and academic affiliation, finding no significant impact of these features (Supplementary Fig. 9). 

Next, we investigate team success as a function of the knowledge diversity and the knowledge broadness of its core members.
In Fig.~\ref{fig:fig3}e, we show the average c5 of the team publications as a function of the knowledge broadness, grouping again cores based on their knowledge diversity.
We observe an inverted U shape relationship between impact and knowledge broadness of the core, for all classes of knowledge diversity, with intermediate values of broadness supporting highest impact. 
When the knowledge broadness of a core is very low, the impact of a team's work might be limited to narrow disciplinary fields. 
Yet, when a core's knowledge base becomes too broad, a team's work yields a lower impact. Similar observations were made at the level of single works, where highest impact is obtained in papers which display a balance between conventional and atypical combinations of prior work \cite{uzzi2013atypical}.
Furthermore, we observe that, for almost any value of knowledge broadness, knowledge diversity has a negative relationship with the average core impact.
All in all, our findings complement previous results on the complex relation between disciplinary diversity and impact in teams\cite{alshebli2018preeminence}.

We conclude our analysis by assessing the role of transient members on team performance. 
We classify core publications into two groups, namely those authored exclusively by core members and those including other contributors, and evaluate the average impact for these two categories. 
Only teams with at least one publication with non-core members and one with only core members are kept for a fair comparison.
Fig.~\ref{fig:fig3}f shows the distribution of the average c5 across scientific cores for the two groups of papers.
The analysis reveals that publications involving transient members generally show lower impact compared to those authored by the core only, suggesting that even though transient members add diversity to the team (see Supplementary Fig.~S4), they might not boost impact.

\section*{Discussion}
In this paper, we introduced the notion of team careers to unravel the determinants and the temporal patterns behind persistent collaborations in science. 
We investigated the features of half a million persistent teams and the patterns governing their formation and lifespans, composition, production and eventually impact. 

Core teams of three scientists were prevalent, highlighting the need to study team careers beyond pairwise collaborations \cite{son2023untangling}. 
Larger cores formed slower, and had shorter career lengths. 
Persistent collaborations with smaller cores often featured a mix of young and established researchers, while larger cores included members from all age groups. 
Additionally, members of persistent cores were affiliated with the same university for the majority of cores. 
A wide range of diverse degrees of disciplinary composition was observed, including teams with large knowledge broadness, likely linked to the rising popularity of interdisciplinary research topics \cite{porter2009science, van2015interdisciplinary}.  
Teams with a higher knowledge diversity were found to have shorter lifespans and longer formation times, highlighting the importance of topic synergy for persistence.

In our examination of temporal patterns in the success of persistent teams, we found that the highest-impact paper can occur at anytime, with the same probability, throughout the teams career. 
Over extended periods, however, we found that the average magnitude of impact is higher for publications in the first half of the team career, in agreement with earlier observations that freshness is associated with high multidisciplinary impact \cite{zeng2021fresh}.   
Besides, hot-streaks were observed in team careers, with the highest-impact papers showing a tendency to be clustered in time, mirroring findings on individual scientific  \cite{liu2018hot,liu2021understanding} as well as artistic careers \cite{williams2019quantifying}. 
 
Previous research shows that a time-aggregated analysis captures multi-university teams to be more successful \cite{jones2008multi}, yet, we find that, for persistent collaborations, this is the case only since the 2000s. Coinciding with the advent of ICT technologies and the internet, remote research teams achieve higher impact. Yet, recent studies indicate that remote work results in fewer bridges and encourages asynchronous communication among employees \cite{yang2022effects}, while also being linked to a lower likelihood of disruptive scientific ideas as compared to onsite collaborations \cite{lin2023remote}. Impactful teams seemed to strike a balance in terms of knowledge broadness, while diversity among team members was associated to low citation accumulation, in accordance with previous analysis at the level of single manuscripts \cite{uzzi2013atypical}, where the highest impact is obtained for papers that display a balance between the conventional and atypical combinations of prior work. 
Contrasting with citation impact, productivity was highest for teams that were topic focused or ones that maximized knowledge breadth. Moreover, publications authored solely by core members yield higher impact compared to those involving transient non-core members, hinting that expanding teams may not always enhance outcomes for persistent collaborations.

While our work reveals features of successful persistent collaborations, whether the observed patterns reflect a causal relationship between team composition and success, or success is simply a prerequisite for collaborations to persist and survive, is an open question which might be clarified in future investigations.
Besides, future works investigating team composition might benefit from integrating additional information about team members, including gender and ethnicity. 
Indeed, we know at the level of teams defined as single set of coauthors that gender diversity among members plays a role in the impact achieved by the publication \cite{yang2022gender}, paving the way for a similar investigation on the role of gender diversity for persistent collaborations. 
We note that bibliometric databases do not include self-declared gender or race by authors and inferring such metadata from names often introduces algorithmic biases \cite{kozlowski2022avoiding} which are distributed unevenly among other demographic traits \cite{lockhart2023name}, highlighting the difficulties in properly carrying out such analysis. 

In the future, our results can be combined with prior theoretical work \cite{guimera2005team,milojevic2010modes, milojevic2014principles} to build a more complete theory of team assembly in science. 
Besides extracting persistent collaborations active over time, we can study team dynamics, how teams evolve and transform themselves and collectively adapt to a scientific ecosystem that constantly evolves in time \cite{galesic2023beyond}. Furthermore, a cross-disciplinary analysis comparing teams in rapidly evolving fields like AI with those in more stable areas like mathematics, can guide tailored team formation strategies. 

Examining the role of funding on team careers can highlight how external support mechanisms can promote meaningful persistent collaborations. 
Indeed, the scientific ecosystem is largely shaped in response to funding. Over time, disciplines differ in terms of funding support they receive, with interdisciplinary research so far achieving lower funding success \cite{bromham2016interdisciplinary}, likely impacting the longevity of such teams. Moreover, grant success is strongly dependent to the collaboration network \cite{chowdhary2023dependency, chowdhary2023funding}, particularly for young researchers, and this may incentivize participation in teams on strategic rather than scientific grounds. From the perspective of individual careers, how persistent collaborations affect they way in which researchers navigate the knowledge landscape is also an intriguing question to address \cite{singh2024charting}. Long-term career outcomes of individuals who participate in high-impact, persistent collaborative experiences can be compared to those who work in more transient or less successful teams. Notably, early-career collaborations with elite scientists predict subsequent career success \cite{li2019early}. Investigating this aspect further could provide insights into how early career researchers can strategically navigate collaborations and switch between teams to enhance their professional growth and impact. 

Overall, our work identifies persistent teams in science and captures temporal as well as compositional patterns of success in shared careers.  Our research informs both scientists in building their collaborations and funders in choosing which research teams to support and promote.

\section*{Methods}
\subsection*{Extracting persistent cores}
In this work we analysed publication data from OpenAlex \cite{openalex,chawla2022massive}. 
This database provides publication metadata, topic classification and citation records for around 205 million
journal papers since 1900, covering 90 million scientists disambiguated using machine learning algorithms and integration with ORCID ids of scientists to identify authors \cite{openalex_authordisambiguation}. We curate all scientists with substantial publication records, keeping those with at least 20 papers, resulting in $4,000,926$ scientists.
In this context, identifying the core members of long-lasting scientific teams corresponds to detecting the maximal sets of significantly co-publishing authors~\cite{musciotto2022identifying}. In fact, these sets represent groups of authors that consistently work together, pruned from the members who only occasionally have published with them. To extract those, we start by constructing the underlying hypergraph of scientific collaborations, i.e. a generalized network that naturally encodes group relationships, usually called hyperedges\cite{battiston2020networks}.
Specifically, we construct a hypergraph whose nodes are authors and whose hyperedges represent joint publications among them.

In order to assess the statistical significance of collective interactions among sets of nodes in a hypergraph, one needs to take into account the heterogeneity of node activity. Indeed, if on one side the easiest way would be setting a fixed threshold on the minimum number of repeated interactions needed to consider a group of nodes a statistically validated set, this approach is sub-optimal, due to the multiscale nature of collaboration networks and the varying activity levels among scientists (the same threshold can be too restrictive for a author with limited publication records and very permissive for a more prolific author). To overcome this limitation, our method is based on a null hypothesis approach, that naturally tunes this threshold on the activity of the involved nodes. For simplicity, we start from the case of 3 nodes $i,j,k$ that co-interact $N_{ijk}$ times in hyperedges of size $n>=3$. The three nodes appear respectively in $N_i,N_j,N_k$ hyperedges.  Under the null hypothesis that each node selects randomly the hyperedges to which it participates - and thus its $n-1$ counterparts in a hyperedge of size $n$ - the probability of observing $i,j,k$ interacting $N_{ijk}$ times is

{\footnotesize
\begin{align*}
		p(N_{ijk}) &= \sum_{X} H(X|N,N_i,N_j)\times  H(N_{ijk}|N,X,N_k) \\
						 &= \frac{1}{\binom{N}{N_j}\binom{N}{N_k}}\sum_X \binom{N_i}{X}\binom{N-N_i}{N_j-X}\binom{X}{N_{ijk}}\binom{N-X}{N_k - N_{ijk}},
					 \stepcounter{equation}\tag{\theequation}\label{eq:prob3}
	\end{align*}
}
where $H(N_{AB}|N,N_A,N_B)$ is the hypergeometric distribution that computes the probability of having an intersection of size $N_{AB}$ between two sets $A$ and $B$ of size $N_A$ and $N_B$ given $N$ total elements. The probability $p(N_{ijk})$ in Eq.~\ref{eq:prob3} represents the probability of having a random intersection of size $N_{ijk}$ between the three sets of hyperedges of nodes $i,j,k$ out of $N$ total hyperedges~\cite{wang2015efficient}, and is obtained through the convolution of two instances of the hypergeometric distribution. Starting from Eq.~\ref{eq:prob3}, we then compute a p-value for the triplet that contains $i$,$j$ and $k$ through the survival function,
\begin{equation}
	\label{eq:pvalue_3}
		p(x\geq N_{ijk}) = 1 - \sum_{x=0}^{N_{ijk} - 1} p(x).
	\end{equation}
The p-value represents the probability of observing $N_{ijk}$ or more hyperedges that contain - but are not limited to - the nodes $i,j,k$. The smaller the p-value, the higher the possibility that $i,j,k$ constitute a significant set of size 3.

In the approach of \cite{musciotto2021detecting,musciotto2022identifying}, N corresponds to the total number of papers in the corpus. In our case, this choice for the value of $N$ implies that all hyperedges are equally accessible by all nodes. It is, however, unrealistic that a scientist could have participated in all papers, due to limited time and resources. Thus, we decided to bound the number of papers a scientist or a group of scientists could have worked on. Specifically, we approximate $N$ by summing the number of unique publications by the scientists and all their coauthors, for each of the members of the group we are testing for significance.

For a generic hyperedge of $n$ nodes, Eq.~\ref{eq:prob3} becomes
{\footnotesize
	\begin{align*}
		p(N_{1...n}) = &\sum_{X_{12}} H(X_{12}|N,N_{1},N_{2})\times \\
							 &\times \sum_{X_{123}} H(X_{123}|N,X_{12},N_{3})\times... \nonumber \\
							 &...\times\sum_{X_{12...n-1}}H(X_{12...n-1}|N,X_{12...n-2},N_{n-1})\times\\
							 &\times H(N_{12...n}|N,X_{12...n-1},N_{n}).
\stepcounter{equation}\tag{\theequation}\label{eq:probn}
\end{align*}
} 

How to set a rigorous criterion to assess whether a group of $n$ scientists is statistically significant, once we have calculated the associated p-value? In order to do so, we test all p-values against a threshold of
statistical significance $\alpha$, after including a multiple hypothesis test correction which is needed because of the high number of tests - one per each group. In all the results presented in this paper we use $\alpha = 0.01$. Coherently with the approach in ~\cite{musciotto2022identifying} we consider all smaller combinations of nodes constituting a significant set to be themselves significant
sets. In other words, if the interaction $i$, $j$ and $k$ is significant, we do not test also the three couple obtained through combination of the triplet.
This means that we start testing from the largest set and we then proceed towards the smallest. If a set of size n passes our statistical test and is thus selected as significant because it rejects the null hypothesis, we do not test any of its smaller subsets. In this way, the obtained statistically significant sets can be considered maximal, i.e. for each of them there is no larger set that includes all its members that is also statistically significant. 

\noindent The extraction of team cores resulted in over half-a-million persistent collaborations, with size ranging from 2 to 10. 
Yet, we limit our analysis to cores of size 2 to 6, as bigger cores are rare and statistics are insufficient for an in-depth analysis. 

\subsection*{Knowledge broadness and knowledge diversity}

\noindent We start from the OpenAlex dataset, where researchers are associated with different scientific concepts with a score varying between 0 and 100. 
The concepts are organized in a hierarchy with 19 root-level concepts representing major disciplines such as physics, chemistry, computer science and so on, and 5 layers of descendants branching out from them, for a total of 65,000 concepts \cite{openalex_concepts}. 
For our analysis, we consider the topmost layer of the concept hierarchy, namely the scientific disciplines. For each member of the team, we evaluate a knowledge vector where each component represents how strongly that scientist is associated with a scientific discipline. Each entry of the vector consists of the topic score normalized by the sum of the scores.

Knowledge broadness captures the breadth of the combined expertise of the persistent core.
To calculate it, we first consider the sum of the knowledge vectors of the core members, normalized so that the components of the combined knowledge vector sum to 1.
We then evaluate the entropy of this combined knowledge vector, calculated as $-\sum_x p_x log_2 p_x$, where $x$ is a scientific discipline and $p_x$ represent how strongly the team is connected to it.
Finally, to obtain the knowledge broadness, the entropy of the vector is normalized by its maximum theoretical value of $log_2(N)$ which corresponds to the case where the team is uniformly spread across all N disciplines.
In particular, as Openalex has 19 disciplines in the top layer of its concept hierarchy here we have $N=19$.
For illustration, consider a core of three members, A, B, and C, whose knowledge vectors are A:[physics: 0.5, chemistry:0.5], B:[physics:0.7, chemistry: 0.1, biology: 0.2], C:[physics:1].
The combined knowledge vector for this team would be: [physics:0.85, chemistry: 0.077, biology: 0.077].
The knowledge broadness of the team is thus $\sim0.18$.
The value of knowledge broadness ranges from 0 to 1. 
A value of 0 indicates a team where all members are associated to a single topic, i.e., a monodisciplinary team, while a value of 1 corresponds to a team in which the joint knowledge vector of the team is uniformly distributed across all possible topics in our data. 

Knowledge diversity, on the other hand, quantifies how much team members are different with regard to the scientific disciplines they are associated with. 
Given the knowledge vectors of the core members, we quantify the knowledge diversity as 1 minus the the average cosine similarity between all pairs of knowledge vectors.
For instance, in the example above, the cosine similarity between members A and B ($\sim 0.77$) is calculated, then B and C ($\sim 0.95$), then C and A ($\sim 0.71$). 
The values are then averaged and subtracted from 1 to obtain the knowledge diversity of the core ($\sim0.19$). 
Knowledge diversity ranges between 0 and 1, where a value of 0 indicates all members have the identical knowledge vectors, i.e., all members are associated with the same strength to the same disciplines, whereas a value of 1 indicates that knowledge vectors of all members are non-overlapping, i.e., they are associated to completely different scientific disciplines.

Knowledge broadness and knowledge diversity cover complementary dimensions of team topic composition, as the first describes how different are the topics associated to the team, while the second captures how the members of the team are diverse among each other.
For instance, a collaboration among scientists working in the same discipline will have low knowledge diversity and low knowledge broadness, while a team where members are associated with the same set of multiple disciplines will have low knowledge diversity but high knowledge broadness.

\section*{Data availability}
The data from \href{openalex.org}{openalex.org} used in this work is openly accessible for download using the API \href{https://api.openalex.org/works}{https://api.openalex.org/works}.

\section*{Code availability}
The code used in this study is available at 
\href{https://github.com/chowdhary-sandeep/sciscicareers}{https://github.com/chowdhary-sandeep/sciscicareers}.

\bibliography{main.bib}
	
\section*{Acknowledgements}
L.G. and F.B. acknowledge support from the Air Force Office of Scientific Research under award number FA8655-22-1-7025. 

\section*{Author contributions statement}
S.C. led the study, processed the data and performed the analysis. F.M. provided methodological insights and curated the data. L.G. provided methodological insights. F.B provided methodological insights, directed and supervised the study. 
S.C, L.G., F.B. substantially contributed to the interpretation of the analysis, all authors revised the manuscript.




\newpage
\renewcommand{\thesection}{S\arabic{section}} 
\renewcommand{\theequation}{S\arabic{equation}}
\renewcommand{\thefigure}{\arabic{figure}}

\newcommand{\eref}[1]{Eq.~(\ref{#1})}
\newcommand{\esref}[2]{Eqs.~(\ref{#1})-(\ref{#2})}
\newcommand{\sref}[1]{Section~\ref{#1}}
\newcommand{\fref}[1]{Fig.~S\ref{#1}}
\newcommand{\fsref}[2]{Figs.~S\ref{#1}-S\ref{#2}}
\newcommand{\tref}[1]{Table~S\ref{#1}}

\def\san#1{{\small\color{blue}\textbf{#1}}}
\def\leo#1{{\small\color{purple}\textbf{#1}}}
\def\adrielsa#1{{\small\color{pink}\textbf{#1}}}
\newcommand{\todo}[1]{{\color{red} (TO DO: {#1})}}
\newcommand{\sandeep}[1]{{\color{red} s \thetodocounter: #1}}
\definecolor{teal}{rgb}{0.0, 0.5, 0.5}

\graphicspath{ {./figures/} }


\setcounter{figure}{0}

\renewcommand\thefigure{S\arabic{figure}}
\begin{center}
{\LARGE Supplementary Information for}\\[0.7cm]
{\Large \textbf{Team careers in science: formation, composition and success of persistent collaborations}}\\[0.5cm]
{\large S. Chowdhary, L. Gallo, F. Musciotto, F. Battiston}\\[0.7cm]
{\small $^*$Corresponding author email: battistonf@ceu.edu}\\[2cm]
\end{center}

\section{Formation time as a function of core composition}
\label{sec:S8}
\begin{figure*}[h]
\centering
\includegraphics[width=1\textwidth,trim={0cm 0cm 0cm 0cm},clip]{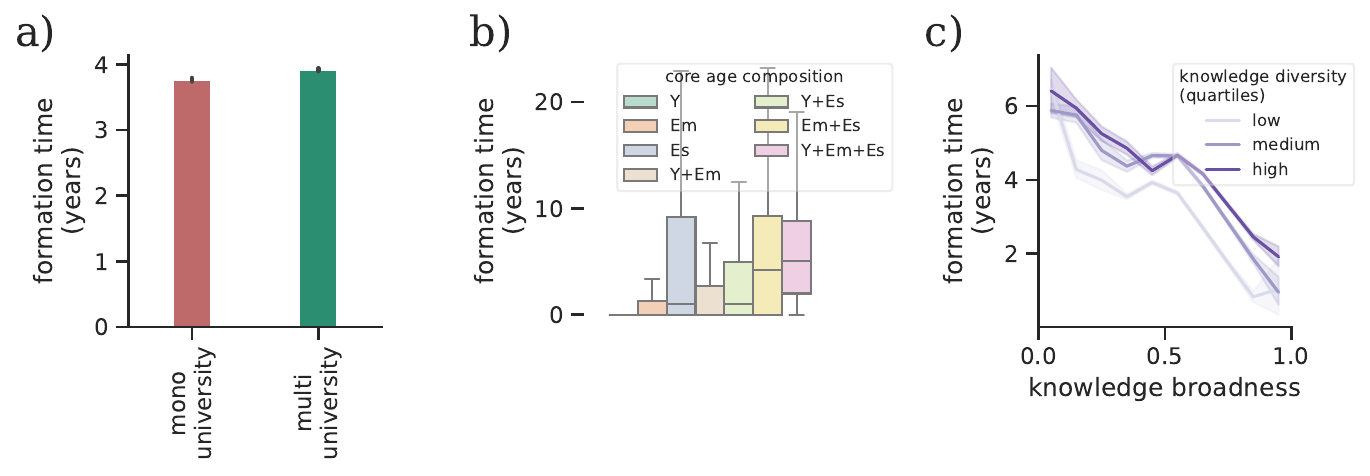}
\caption{\small Formation time of a time as a function of a) diversity in the academic affiliation, b) age composition, and c) knowledge broadness and knowledge diversity.}
    \label{fig:SI8}

\end{figure*}

Extending our investigation of core formation in main text,  we separate cores by academic affiliation, age and disciplinary diversity and measure how the formation time is influenced by these features. 
In terms of affiliation, we find that mono-university teams form faster compared to multi-university ones (panel a of Fig.\ref{fig:SI8}). 
Age is also related to the formation time, with older cores taking longer than younger cores (panel b of Fig.\ref{fig:SI8}). 
A monotonic decrease in formation time is observed as a function of broadness, with cores composed of highly-interdisciplinary scientists assembling fastest (panel c of Fig.\ref{fig:SI8}).

\section{Career length as a function of core composition}
\label{sec:S5}
\begin{figure*}[h]
\includegraphics[width=1\textwidth,trim={0cm 0cm 0cm 0cm},clip]{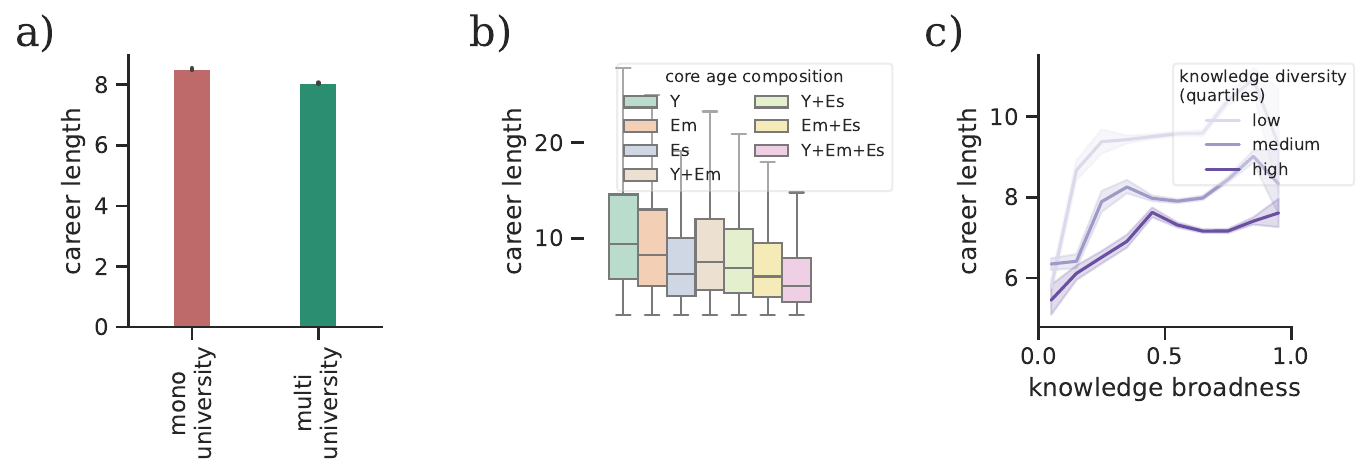}
\caption{\small Career length as a function of a) diversity in the academic affiliation, b) age composition, and c) knowledge broadness and knowledge diversity.}
    \label{fig:SI5}

\end{figure*}

In the main text, we quantified the typical career length of persistent collaborations as a function of team size. 
Here we develop the analysis further and examine how the core's composition determines its lifespan. 
We observe that mono-university cores have slightly longer lifespans compared to those spread across multiple universities (panel a of Fig.\ref{fig:SI5}). Dividing cores by age compositions, we observe 
that teams where members come from the same age group usually have longer careers than cores with diverse ages.
In general, we note that younger teams tend to last longer than older ones (panel b of Fig.\ref{fig:SI5}).
Results on knowledge diversity of cores show that the lifespan of a core is positively correlated with its knowledge broadness and negatively associated with knowledge diversity (panel b of Fig.\ref{fig:SI5}).

\section{Core exclusivity as a function of core composition}
\label{sec:S6}

\begin{figure*}[h]
\includegraphics[width=1\textwidth,trim={0cm 0cm 0cm 0cm},clip]{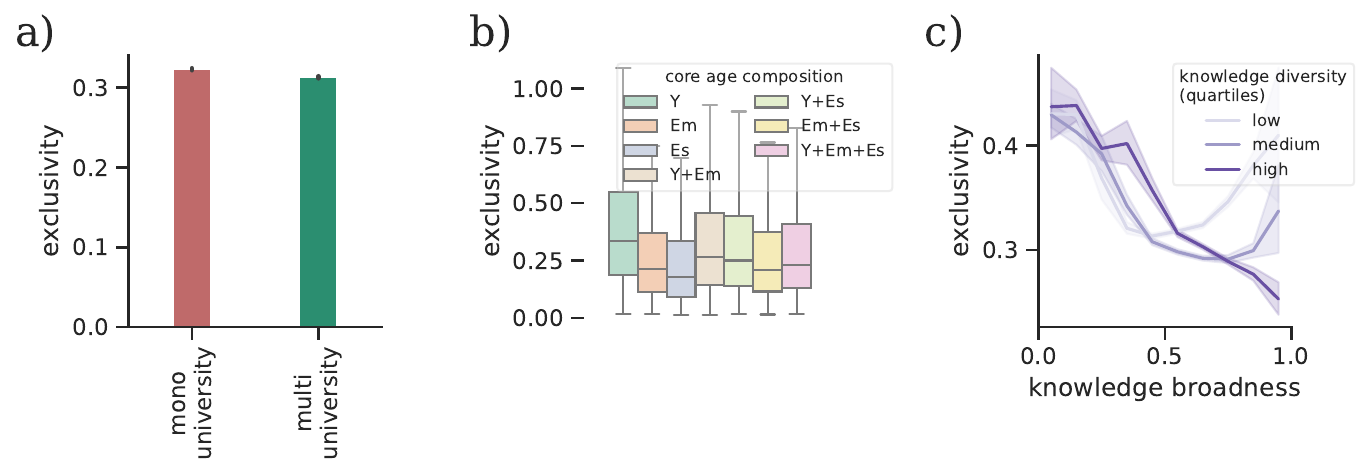}
\caption{\small Core exclusivity as a function of a) diversity in the academic affiliation, b) age composition, and c) knowledge broadness and knowledge diversity.}
    \label{fig:SI6}

\end{figure*}

The members of a persistent core may work exclusively together or explore collaborations outside the core.
We define the exclusivity of a persistent team as the ratio of the number of publications authored by all core members to the total numbers of papers featuring at least one core member.
In general, core tend to be highly exclusive, as more than 30$\%$ of all papers considered are published within a persistent collaboration.
In terms of team composition, we don't observe a significant difference between mono and multi-university cores (panel a of Fig.\ref{fig:SI6}).
Instead, we note that cores made only of young researchers are in general more exclusive (panel b of Fig.\ref{fig:SI6}).
Also, we find that exclusivity decreases as a function of the knowledge broadness, while it is not strongly correlated with knowledge diversity (panel c of Fig.\ref{fig:SI6}).

\section{Contribution of transient, non-core members of teams}
\label{sec:S8}
\begin{figure*}[h]
\includegraphics[width=1\textwidth,trim={0cm 0cm 0cm 0cm},clip]{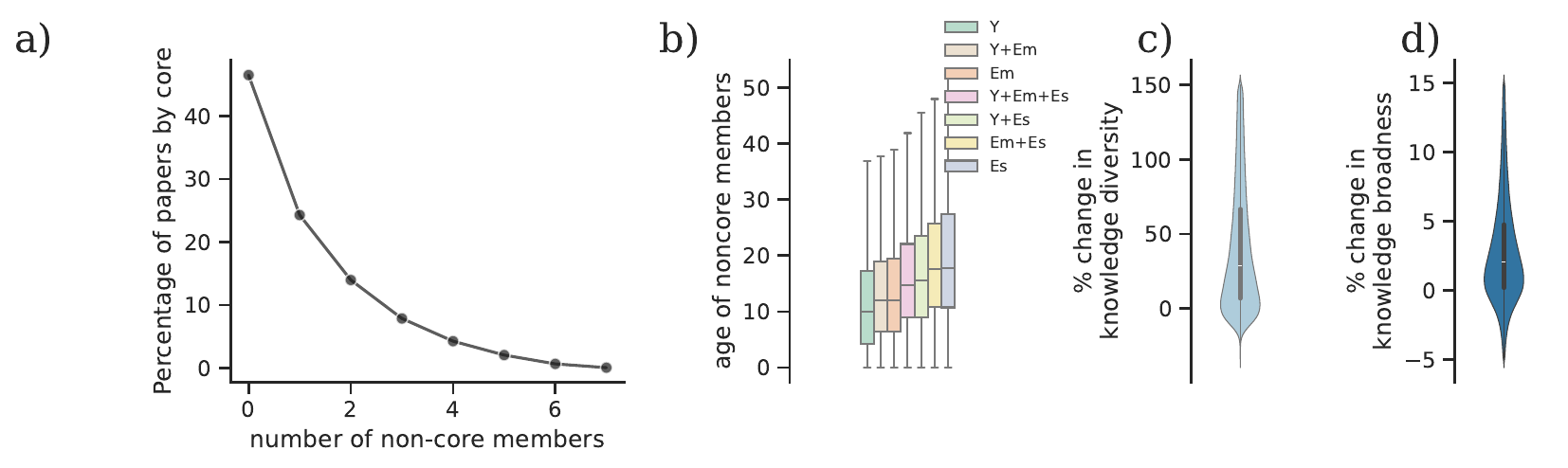}
\caption{\small \textbf{Contribution of transient members of persistent teams.} a) Percentage of papers published as a function of the additional number of non-core members. b) Age distributions of the non-core members as a function of the age composition of the core. c) Percentage change in the team diversity upon addition of non-core members. d) Percentage change in the team knowledge broadness upon addition of non-core members.}
\label{fig:SI9}
\end{figure*}

In the main text, we focused only on the core members of scientific teams. 
In general, however, a team is made by core members that persistently work together and transient members that sporadically publish with the core.
Here, we expand our analysis of teams to this latter group of members.
First, we evaluate the fraction of papers published by persistent collaborations that include a given number of transient members (panel a of Fig.\ref{fig:SI9}). 
We find that over $46.5\%$ of papers are authored exclusively by core members, while $24.3\%$ ($14.0\%$) include 1 (2) non-core members, and less than $20\%$ feature 3 or more transient members.

When categorizing cores by age composition and comparing this to the ages of non-core members (Fig.\ref{fig:SI9}b), we find that cores of young scientists tend to collaborate with younger transient researchers, whereas more established cores often work with older non-core members.

Incorporating non-core members into teams generally enhances knowledge diversity ($+87.7\%$ on average) and knowledge broadness ($+76.7\%$ on average), indicating that transient members add to persistent collaborations a different expertise (panels c and d of Fig.\ref{fig:SI9}). 
However, for $12.3\%$ of cores, knowledge diversity decreases with the addition of a transient member, suggesting potential redundancy, which warrants further investigation. 
Finally, in terms of affiliations, in $95.0\%$ cases non-core members are based in the same university or institute.

\section{Team impact as a function of career stage}
\label{sec:S1}

\begin{figure*}[h]
\centering
\includegraphics[width=.4\textwidth,trim={0cm 0cm 0cm 0cm},clip]{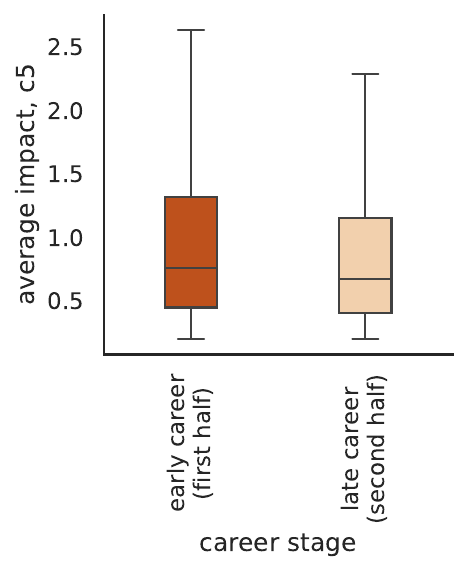}
\caption{\small\textbf{Team impact as a function of career stage.} \small Boxplots show the distribution of the the average impact (c5) of the core papers in the first and second half of their careers.
}
    \label{fig:SI1}

\end{figure*}

\noindent
In the evolution of a team's career, freshness decreases. 
This can have consequences for the success of the team and the impact it gathers. 
To test this, we split the papers of each team (ordered chronologically) into two halves and compute their average impact, measured as the normalized number of citations after five years from the publication. 
Fig.\ref{fig:SI1} shows the distribution of the average impact of the first and second half of the cores' careers.
We observe that persistent collaborations are generally more impactful at in the first half of their careers, in agreement with previous results on the relationship between team freshness and impact.

\section{Hot-streaks in team careers}
\label{sec:S2}

\begin{figure*}[h]
\includegraphics[width=1\textwidth,trim={0cm 0cm 0cm 0cm},clip]{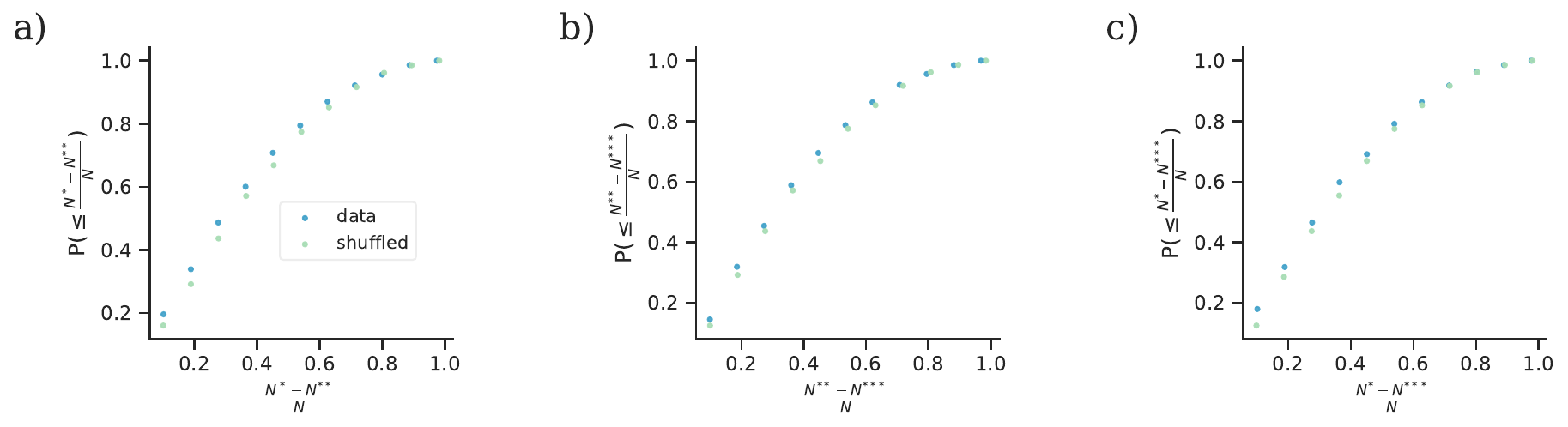}
\caption{\small\textbf{Hot-streaks in team careers.}  (a) Cumulative distributions P($\leq$(N*-N**)/N), where N*-N** denotes the distances in temporal order of the highest and the second-highest impact papers in a core’s career respectively. (b) P($\leq$(N**-N***)/N) where N*-N*** where N*-N** denotes the distances in order of the second-highest and third-highest impact papers.  (c) P($\leq$(N*-N***)/N) where N**-N*** denotes the distances in order of the highest and third-highest impact papers. The shuffled model randomizes the order of impact within a teams career, in all three panels. }
    \label{fig:SI2}

\end{figure*}

It is known that the most-cited papers in an individual scientific careers tend to be clustered in time, a phenomenon known as hot-streak.
To examine the possibility of hot-streaks in team careers, we follow the same approach used for individual careers of scientists. 
We identify the position of the 3 most impactful papers in a team career, N*-th, N**-th and N***-th, respectively, and compute how each pair of them is distant in temporal order, i.e., (N*-N**)/N, (N*-N***)/N and (N**-N***)/N, where N is the total number of publication of the core. 
The normalization by the length of the core's career allows for a fair aggregation over all cores. 
We show the cumulative distributions for the three distances in Fig.\ref{fig:SI2}. 
The figure also shows the distances obtained for a null model that randomizes the order of the papers within a teams career. 
The comparison with the null model indicate that highest impact works in teams' careers are likely to be close in time, namely to occur in bursts, highlighting the presence of hot-streaks in team careers. 

\section{Team age compositions and impact}
\label{sec:S3}
\begin{figure*}[h]
\centering
\includegraphics[width=.5\textwidth,trim={0cm 0cm 0cm 0cm},clip]{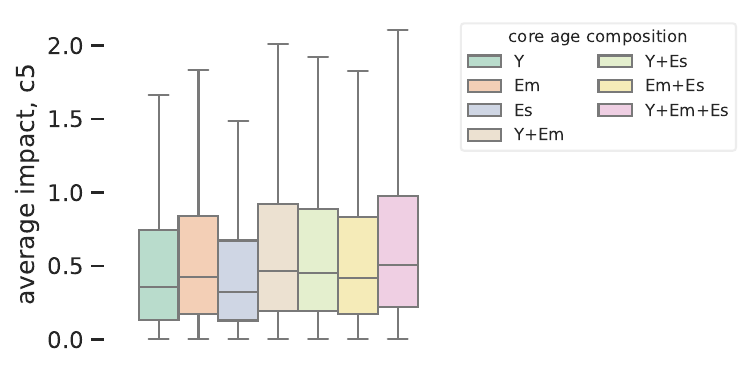}
\caption{\small \textbf{Team age composition and impact.}  Cores are separated based on age composition of members and the distribution of the average impact per paper for each age composition is shown. }
    \label{fig:SI3_5}

\end{figure*}

To understand the role of members' ages on team impact, we measure the average paper impact, i.e., the average normalized number of citations after five years, for all possible core age compositions in our data.
As shown in 
(Fig.\ref{fig:SI3_5}), cores composed of members from the same age groups perform on average worse than collaborations with scientists from multiple age groups.
In particular, we observe that cores whose members come from all age groups, i.e., young, emerging, and established, are those with the highest impact.
Among mono-age group collaborations, we observe that core with emerging authors only tend to outperform those comprised of young scientists only, as well as those where members are all established researchers. 
This result calls for further studies on how teams with similar or diverse academic ages communicate and organize. 

\section{Time aggregated analysis of affiliation diversity vs impact}
\label{sec:S3_5}
\begin{figure*}[h]
\centering
\includegraphics[width=.35\textwidth,trim={0cm 0cm 0cm 0cm},clip]{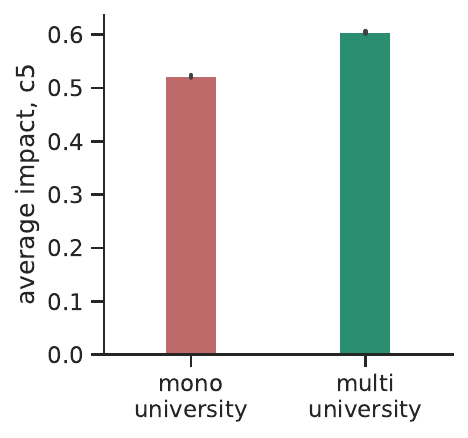}
\caption{\small \textbf{Team impact and geography.}  Cores are separated into those co-located at the same university, and others which multiple institutions and average impact is shown for both categories of cores. Static time-aggregated results show higher impact for multiuniversity cores, matching the observation of Jones et al.}
    \label{fig:SI3}

\end{figure*}

In the main text, we showed that the effects of core's affiliation diversity on team impact has varied throughout the decades, and only since the 2000s have multi-university cores started to outperform mono-university cores in impact.  A time aggregated analysis of our data reveals that average impact for multi-university is indeed higher on average than mono-university teams (Fig.\ref{fig:SI3}).
A possible reason for their observation is the rise in the number of multi-university teams in the last decades which are over-represented in data and overshadow the underwhelming performance of such cores in the past. 
Thus dominating a static (non-time varying) analysis.

\section{Productivity as a function of core composition}
\begin{figure*}[h]
\centering
\includegraphics[width=.7\textwidth,trim={0cm 0cm 8cm 0cm},clip]{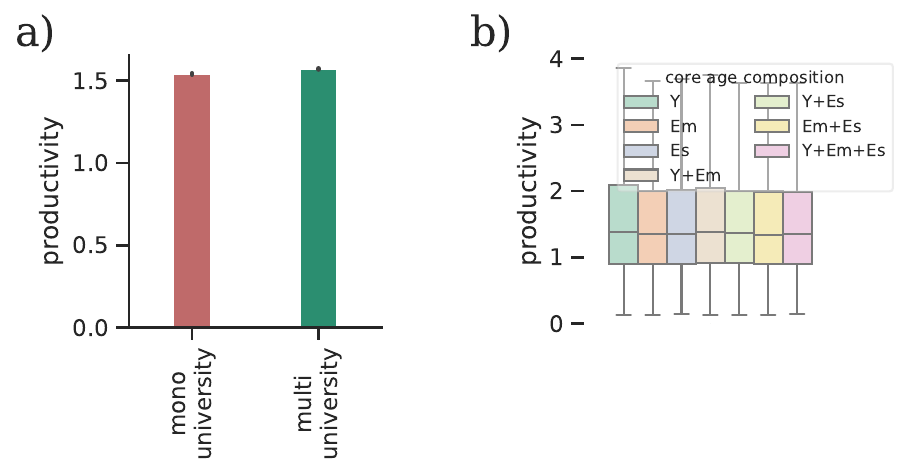}
\caption{\small Productivity as a function of a) core's affiliation and b)  composition. }
    \label{fig:SI4}

\end{figure*}

In addition to impact, measured as the number of citations, another dimension of performance is productivity. 
Here we define productivity as the number of papers published by a core divided by the number of years the core persisted. 
We study the effect of geographical and age composition of core team on the productivity (Fig.\ref{fig:SI4}). 
Our results show no discernible effect of age composition and geographical diversity on the team productivity. 
On the other hand, core's knowledge broadness and diversity were shown to affect productivity (main text).

\label{sec:S4}

\end{document}